\documentclass{jetpl}
\twocolumn

\usepackage{amsmath,amssymb,epsfig,color,pstricks,bm,graphics,alltt}

\begin{document}

\lat

\title{Anion height dependence of T$_c$ and density of states in 
iron based superconductors}

\rtitle{Anion height dependence of T$_c$}

\sodtitle{Anion height dependence of T$_c$ and density of states in Fe -- based superconductors}

\author{E.\ Z.\ Kuchinskii\thanks{E-mail: kuchinsk@iep.uran.ru}, I.\ A.\ Nekrasov\thanks{E-mail: nekrasov@iep.uran.ru}, 
M.\ V.\ Sadovskii\thanks{E-mail: sadovski@iep.uran.ru}}

\rauthor{E.\ Z.\ Kuchinskii, I.\ A.\ Nekrasov, M.\ V.\ Sadovskii}

\sodauthor{Kuchinskii, Nekrasov, Sadovskii }

\sodauthor{Kuchinskii, Nekrasov, Sadovskii}

\address{Institute for Electrophysics, Russian Academy of Sciences, 
Ural Branch, Amundsen str. 106,  Ekaterinburg, 620016, Russia}

\dates{April 2010}{*}

\abstract{
Systematic {\it ab initio} LDA calculations were performed for all the typical
representatives of recently discovered class of iron based high-temperature 
superconductors: REOFe(As,P) (RE=La,Ce,Nd,Sm,Tb), Ba$_2$Fe$_2$As, (Sr,Ca)FFeAs, 
Sr$_4$Sc$_2$O$_6$Fe$_2$P$_2$,  LiFeAs and Fe(Se,Te). Non-monotonic behavior of total 
density of states at the Fermi level is observed as a function of anion height 
relative to Fe layer with maximum at about $\Delta z_a\sim$1.37\AA, attributed
to changing Fe - As (P,Se,Te) hybridization. This leads to a similar 
dependence of superconducting transition temperature T$_c$ as observed in the 
experiments. The fit of this dependence to elementary BCS theory produces 
semiquantitative agreement with experimental data for T$_c$ for the whole class
of iron based superconductors. The similar fit to Allen -- Dynes formula
underestimates T$_c$ in the vicinity of the maximum, signifying the possible
importance of non -- phonon pairing in this region. These results unambiguously
demonstrate that the main effect of T$_c$ variation between different types
of iron based superconductors is due to the corresponding variation of the
density of states at the Fermi level. 
}

\maketitle

Recent discovery of the new class of iron based high-temperature superconductors 
\cite{kamihara_08} has ignited almost unprecedented stream of experimental and 
theoretical studies (for the review of an early work see \cite{UFN_90,Hoso_09}). 
Despite the immense progress in understanding of these systems, the nature 
(mechanism) of superconducting pairing, as well as the reasons for high values 
of superconducting temperature $T_c$, are still under debate. In a recent 
preprint \cite{hPn2} Mizuguhci et al. have established an interesting anion 
height dependence of $T_c$ for the typical representatives of Fe - based 
superconductors, demonstrating  almost symmetric curve with a peak around 
$\Delta z_a\sim$ 1.37\AA. Below we present an explanation of this dependence, 
attributing it to the effect of the appropriate variation of the total density 
of states at the Fermi level  within standard BCS-like approach.

The main representatives of the class of iron (pnictides, chalcogenides) based 
superconductors known at the moment are:

\begin{enumerate}
\item{Doped RE111 (RE=La,Ce,Pr,Nd,Sm,Tb,Dy) with  T$_c$ about 25--55 K, with 
most typical representatives such as LaO$_{1-x}$F$_x$FeAs 
\cite{kamihara_08,chen,zhu,mand,chen_3790,chen_3603,ren_4234,ren_4283,TbDy} and 
LaO$_{1-x}$F$_x$FeP~\cite{laofep} with much lower T$_c$=6.6~K.}
\item{Doped A122 (A=Ba,Sr), such as Ba$_{1-x}$K$_x$Fe$_2$As$_2$ 
\cite{rott,ChenLi,Chu,Bud} and T$_c$ about 38 K.}
\item{Li$_{1-x}$FeAs with T$_c$=18 K~\cite{cryst,wang_4688}.}
\item{(Sr,Ca,Eu)FFeAs \cite{Tegel,Han} with T$_c$=36 K \cite{Zhu}.}
\item{Sr$_4$(Sc,V)$_2$O$_6$Fe$_2$(P,As)$_2$ with T$_c$=17 K \cite{42622}.}
\item{FeSe$_x$, FeSe$_{1-x}$Te$_x$ with T$_c$ up to 14 K \cite{FeSe}.}
\end{enumerate}

There is now a plenty of papers on LDA (local density approximation) calculation
of the band structure of  La111 \cite{singh,dolg,mazin}, LaOFeP \cite{lebegue},
RE111 series \cite{Nekr}, BaFe$_2$As$_2$ \cite{Nekr2,Shein, Krell},
LiFeAs \cite{Nekr3,Shein2}, (Sr,Ca)FFeAs \cite{Nekr4,Shein3}, Sr42622 
\cite{Shein4} and Fe(S,Se,Te) \cite{SinghFeSe}. Below we present some of the results of 
our continued work on LDA electronic structure, along the lines of 
Refs.~\cite{Nekr,Nekr2,Nekr3,Nekr4}, covering all typical representatives of the 
whole class of iron based superconductors and taking into account some new 
structural data.

Iron based high-T$_c$ superconductors in general have tetragonal structure with the space group $P$4/$nmm$
(RE1111, LiFeAs, Sr42622, Fe(Te,Se), SrFFeAs) and the space group $I$4/$mmm$ (Ba122).
For the $P$4/$nmm$ systems Fe ions occupy positions (2b)(0.75, 0.25, 0.5),
and anion ions A=P,As,Se,Te -- (2c) (0.25,0.25,$z_a$) and for Ba122 Fe(4d) (0.5,0,0.25), 
As(4e)(0,0,$z_a$). Corresponding experimental lattice parameters and atomic 
coordinates (used in this work) are collected in Table~1.

Physically important electronic bands (those which cross the Fermi level)
are formed by antibonding Fe(3d)-A(p) states of FeA$_4$ tetrahedron layer.
Here A denotes different types of anions: P,As,Se,Te.
To calculate electronic structure of compounds listed in Table~1
linearized muffin-tin orbitals method (LMTO)~\cite{LMTO} with default settings
was employed (except for Re111 systems, where Re-4f states were taken as a pseudocore states).
Obtained results are in good agreement with other LDA calculations
by other authors.

Motivated by the results of Ref. \cite{hPn2} we present here our LDA calculated 
total density of states N$(E_F)$ as a function of anion height $\Delta z_a$ with 
respect to Fe layer. Corresponding dependence is plotted in Fig.~1 with circles.
We can see that N$(E_F)$ has an interesting behavior with clear maximum at 
about $\Delta z_a\sim$1.37\AA\  (see also Table~2). 
Such nonmonotonous behavior can be explained by hybridization effects.
Namely, as a governing structural parameter characterizing hybridization 
strength one can chose $a$-Fe-$a$ angle --  an angle between anions ($a$) and Fe 
within the same tetrahedron. The value of the angle corresponding to the  
strongest hybridization is 109.45$^\circ$, i.e.  for an ideal anion tetrahedron 
with Fe in the very center of it.
Other crystal structure parameters which might be marked as importand here such as
Fe-Fe, Fe-$a$ or $a$-$a$ distances are not changed very much from system to 
system and do not have any transparent dependence of $\Delta z_a$. The values of 
these distances are about following 2.8\AA, 2.4\AA~and 3.85\AA~with slight 
lowering for LaOFeP, LiFeAs and Fe(Te,Se) compounds.

From Table~1 one can see that compounds with highest N$(E_F)$ values have the 
$a$-Fe-$a$ angle very close to this value. Decrease or increase of this angle 
leads to N$(E_F)$ drop from this maximum value. This comes from partial DOS
behavior. The strongest hybridization corresponds to the strongest 
bonding-antibonding splitting. Since antibonding band DOS grows monotonically 
with binding energy~\cite{Nekr,Nekr2,Nekr3,Nekr4} stronger
hybridization will lead to higher values of N$(E_F)$. With lowering of 
hybridization bonding-antibonding splitting goes down together with N$(E_F)$.

\begin{figure}[htb]
\includegraphics[clip=true,width=0.5\textwidth]{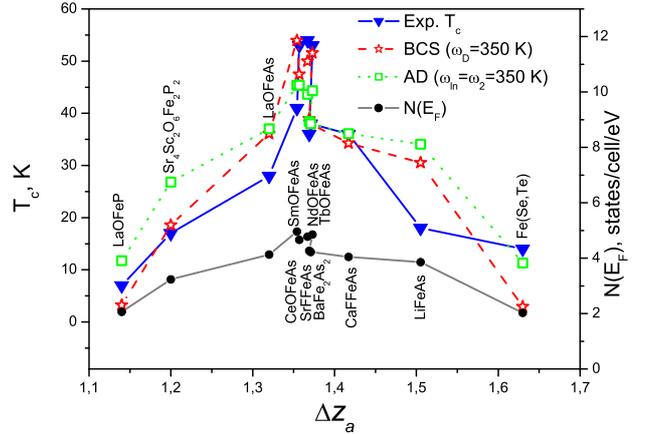}
\label{fig1}
\caption{Fig. 1. LDA calculated total DOS values $N(E_F)$ (circles, right scale) 
and superconducting transition temperatures $T_c$ (left scale) obtained from 
simple BCS (stars) and Allen-Dynes expressions (squares), as well as 
experimental $T_c$ values (triangles) versus anion height $\Delta z_a$ over Fe 
layer for a number of iron based high temperature superconductors.} 
\end{figure}

The $\Delta z_a$ dependence of N$(E_F)$ inevitably leads to the corresponding 
dependence of superconducting critical temperature $T_c$. To estimate this we, 
first of all, use the elementary BCS expression: 
$T_c=1.14\omega_D e^{-1/\lambda}$, where $\omega_D$ is the characteristic 
frequency of collective excitations involved in pairing interaction (phonons, 
spin fluctuations etc.), and $\lambda=gN(E_F)/2$ is the dimensionless
pairing interaction constant ($g$ is the appropriate dimensional coupling 
constant). In the following we take $\omega_D$=350 K in rough accord with 
neutron scattering experiments on phonon density of states for La111 
\cite{Christianson} and Ba122\cite{Mittal} systems. We fix $g$ to fit the 
experimental value of $T_c$ for Ba122 system since this system possesses 
probably most stable value of $T_c$ (about 38 K) with respect to the way of 
sample preparation and doping. Thus we obtain the value of dimensionless 
coupling constant $\lambda$=0.43. Then just fixing the value of $g$ as for Ba122 
we obtain $T_c$ values for all other systems, taking into account the 
appropriate change of the density of states. 
Rather surprisingly we observe almost quantitative agreement with experimental 
data on $T_c$ (see triangles on Fig.~1 and Table~2). Note that we can even obtain the right 
order of $T_c$ values for 1111 systems with different rare-earth elements as due 
to rather small difference of corresponding densities of states, which
were not obtained in our previous work \cite{Nekr}, where we just fixed 
$\Delta z_a$ to the only known at that time experimental value for LaOFeAs. 
However, the calculated value of $T_c$ for LaOFeAs system is still rather higher 
than most typical experimental value of 26-28K. At the same time, the samples of 
this system obtained via high pressure synthesis \cite{Lu41} demonstrated much 
higher values of $T_c\sim$ 41 K, which is pretty closer to our calculated values. 
Also the notable deviation of our calculated $T_c$ for LiFeAs system may be 
attributed both to the crudeness of our model (e.g. our use of a single value of
$\omega_D$ for all compounds), as well as to probable experimental uncertainties 
of $T_c$ in this system. 

In principle, for the number of systems under consideration we can obtain even 
better results if we use the multiple band BCS-like approach, along the lines 
of Ref. \cite{KS09}. However, to reduce the number of free
parameters, the multiple band model fit requires additional information on the 
relations between energy gaps on different Fermi surface sheets (cylinders), 
which at present is only availbale for some of 122 systems.

It is well known that the elementary BCS-like expression for $T_c$ has a 
tendency to overestimate the role of the density of states at the Fermi level. 
As an alternative we try the same approach estimating superconducting 
critical temperature $T_c$ using Allen -- Dynes interpolation formula 
(which is probably the best semi-analytic expression for $T_c$ in case of 
electron-phonon pairing mechanism, including the strong coupling region) 
\cite{AD}:
\begin{equation}
T_c=\frac{f_1f_2\omega_{ln}}{1.20}\exp\left(-\frac{1.04(1+\lambda)}
{\lambda-\mu^*-0.62\lambda\mu^*}\right),
\label{AllenDynes}
\end{equation}
where
\begin{eqnarray}
&& f_1=[1+(\lambda/\Lambda_1)^{3/2}]^{1/3},\qquad
\Lambda_1=2.46(1+3.8\mu^*), \nonumber\\
&& f_2=1+\frac{(\bar\omega_2/\omega_{ln}-1)\lambda^2}{\lambda^2+\Lambda^2_{2}},\
\Lambda_2=1.82(1+6.3\mu^*)(\bar\omega_2/\omega_{ln}),\nonumber
\end{eqnarray}
and $\bar\omega_2=<\omega^2>^{1/2}$, $\omega_{ln}$ are square root average and 
average logarithm of phonon frequency. Assuming for simplicity 
$\omega_{ln}\approx\bar\omega_2$=350~K, and taking the optimistic value of Coulomb 
pseudopotential $\mu^*=0$, we repeat our previous analysis, fixing first 
$\lambda$=0.97 for Ba122 and then changing only the density of states as obtained in 
our calculations for all other systems. The results for $T_c$ obtained in this 
way from Allen -- Dynes expression (\ref{AllenDynes}) are shown in Fig.~1 by
squares (see also Table~2).

We can see that Allen -- Dynes expression produces a kind of a lower bound $T_c$ 
estimate, with obvious deficit in $T_c$ values in the vicinity of maximum. This 
deficit may signify the importance of non phonon pairing mechanism to obtain 
maximal values of $T_c$ in FeAs superconductors. However, our main conclusion
on important correlation of $T_c$ with the values of the density of states at 
the Fermi level remains intact.

In fact we do not adhere at the moment to any specific pairing mechanism. 
Main objection to electron-phonon pairing in iron based superconductors comes
from microscopic calculations, e.g. those of Ref. \cite{dolg}. At the same time,
there are experiments on isotope effect \cite{iso1,iso2}, which support the
importance of this mechanism, though the other isotope experiments \cite{iso3}
produce quite opposite picture. 

Our choice of characteristic phonon frequencies in the preexponential factor of 
BCS and Allen -- Dynes expressions for $T_c$ is used only as a kind of an 
estimate. What is important to us, is  the well known fact that the 
dimensionless pairing constant is proportional to the total density of states 
in almost any BCS-like model of superconducting pairing, with some additional  
modifications in the case of multiple band models \cite{KS09}.

In conclusion, our results show unambiguous correlation of the values of
superconducting $T_c$ and those of the total density of electronic states at
the Fermi level for the whole class of iron based superconductors, thus
supporting the usual BCS-like pairing mechanism in these systems.

This work is partly supported by RFBR grant 08-02-00021 and was performed
within the framework of programs of fundamental research of the Russian Academy
of Sciences (RAS) ``Quantum physics of condensed matter'' (09-$\Pi$-2-1009) and 
of the Physics Division of RAS  ``Strongly correlated electrons in solid states'' 
(09-T-2-1011). IN thanks Grant of President of Russia MK-614.2009.2,
interdisciplinary UB-SB RAS project, and Russian Science Support
Foundation.

\onecolumn

\begin{table}[htb]
\center
\footnotesize
\caption{Table 1. Experimental crystallographic data for iron based superconductors.}
\label{tab1}
\begin{tabular}{|l|c|c|c|c|c|c|}
\hline
System   &  $\Delta z_a$, \AA & $a$, \AA &  $c$, \AA  & $z_{Re}$ & $z_{a}$ & $\angle_{a-Fe-a}$ \\
         &                    &          &            &          &         &                   \\
\hline
LaOFeP                        & 1.140 & 3.9636 & 8.5122 & 0.1487 & 0.6339 & 104.4   \\
Sr$_4$Sc$_2$O$_6$Fe$_2$P$_2$  & 1.200 & 4.0160 & 15.543 & --     & 0.5772 & 105.2   \\
LaOFeAs                       & 1.320 & 4.0353 & 8.7409 & 0.1415 & 0.6512 & 107.5   \\
CeOFeAs                       & 1.354 & 3.9959 & 8.6522 & 0.1480 & 0.6565 & 108.4   \\
SmOFeAs                       & 1.357 & 3.9270 & 8.4413 & 0.1420 & 0.6608 & 108.8   \\
NdOFeAs                       & 1.367 & 3.9476 & 8.5446 & 0.1440 & 0.6600 & 110.5   \\
TbOFeAs                       & 1.373 & 3.8530 & 8.2990 & 0.1447 & 0.6654 & 109.7   \\
SrFFeAs                       & 1.369 & 4.0110 & 8.9650 & 0.1598 & 0.6527 & 108.6   \\
BaFe$_2$As$_2$                & 1.371 & 3.9090 &13.2122 & --     & 0.3538 & 109.3   \\
CaFFeAs                       & 1.417 & 3.8780 & 8.5920 & 0.1505 & 0.6649 & 110.4   \\
LiFeAs                        & 1.505 & 3.7914 & 6.3642 & 0.8459 & 0.2635 & 112.7   \\
Fe(Se,Te)                     & 1.630 & 3.8215 & 6.2695 & --     & 0.2599 & 111.5   \\

\hline
\end{tabular}
\end{table}
\normalsize

\begin{table}[htb]
\center
\footnotesize
\caption{Table 2. LDA total DOS N(E$_F$), calculated and experimental T$_c$ for iron based superconductors.}
\label{tab2}
\begin{tabular}{|l|c|c|c|c|c|}
\hline
System                        &$\Delta z_a$, \AA & N(E$_F$),      & T$^{BCS}_c$, K &  T$^{AD}_c$, K  & T$_c^{Exp}$, K\\
                              &                  & states/cell/eV &                &                 &               \\
\hline
LaOFeP                        & 1.140            & 2.06           &    3.2         &    12           & 6.6 \\
Sr$_4$Sc$_2$O$_6$Fe$_2$P$_2$  & 1.200            & 3.24           &    19          &    27           & 17  \\
LaOFeAs                       & 1.320            & 4.13           &    36          &    37           & 28  \\
CeOFeAs                       & 1.354            & 4.96           &    54          &    43           & 41  \\
SmOFeAs                       & 1.357            & 4.66           &    48          &    37           & 53  \\
NdOFeAs                       & 1.367            & 4.78           &    50          &    44           & 54  \\
TbOFeAs                       & 1.373            & 4.85           &    52          &    45           & 53  \\
SrFFeAs                       & 1.369            & 4.26           &    38          &    39           & 36  \\
BaFe$_2$As$_2$                & 1.371            & 4.22           &    38          &    38           & 38  \\
CaFFeAs                       & 1.417            & 4.04           &    34          &    36           & 36  \\
LiFeAs                        & 1.505            & 3.86           &    31          &    34           & 18  \\
Fe(Se,Te)                     & 1.630            & 2.02           &     3          &    11           & 14  \\

\hline
\end{tabular}
\end{table}
\normalsize

\end{document}